\documentstyle[twocolumn,prc,aps]{revtex}
\begin{document}
\bibliographystyle{unsrt}
\draft
\title{Functional approach to the electromagnetic response function:\\
the Longitudinal Channel
}
\author{R.Cenni and P.Saracco}
\address{Dipartimento di Fisica dell'Universit\`a di Genova\\
and I.N.F.N. - Sez. Genova, Via Dodecaneso,33 - 16146 Genova
(Italy)}
\date{\today}
\maketitle
\begin{abstract}
In this paper we address the (charge) longitudinal electromagnetic response
for a homogeneous system of nucleons interacting via meson exchanges in
the functional framework. This approach warrants consistency if the
calculation is carried on order-by-order in the mesonic loop expansion with
RPA-dressed mesonic propagators. At the 1-loop order and considering $\pi$,
$\rho$ and $\omega$ exchanges we obtain a quenching of the response, in
line with the experimental results.
\end{abstract}

\section{Introduction}

The inclusive cross section for electron scattering off a nuclear
target reads
\begin{eqnarray}
\frac{\displaystyle d\sigma}{\displaystyle dE^\prime d\Omega}
&=&\sigma_M
\left\{
\frac{\displaystyle Q^4}{\displaystyle |\bf q|^4}R_L(\omega,Q^2)
\right.\nonumber\\ &&\quad +\left.
\left[\frac{\displaystyle Q^2}
{\displaystyle 2|\bf q|^2}+\tan^2\frac{\theta }{2}\right]
R_T(\omega,Q^2)
\right\}
\end{eqnarray}
and allows the experimental separation of the
longitudinal and transverse electromagnetic response functions
$R_{L/T}$ by means of a Rosenbluth plot of data taken at the
same $Q^2$ and different angles.

Such separation is grounded on two approximations, namely the
one--photon--exchange and the negligibility of
Coulomb distortion of the electrons. In the inclusive cross section
the two photons exchange corrections are of order
of 1\%\cite{FriRo-74} and may be safely neglected. Instead,
the use of different kinematics, differently  affected by the
Coulomb distortion, could make the
separation somehow questionable\cite{CoHe87,Tra88,TraTuZg88}.

Even within these limits, however, the Rosenbluth separation of
$R_{L/T}$ has recently become available for a wide variety of
nuclei, in different energy and momentum regions
\cite{Al80,Ba83,Me84,Me85,Ma85,De86,Bl86,Qu88,YatWi93} and
has provided new, unexpected outcomes: the central problem
seems to be that a Free Fermi Gas (FFG) model is roughly able to
reproduce the transverse response -- 2p-2h corrections improve
the agreement with the experimental data\cite{AlErMo-84} within
a mesonic theory; on the contrary, the FFG model prevision in the
longitudinal channel is almost twice as large with respect to the
experimental results. The most recent data on $^{40}Ca$
\cite{YatWi93} seem to weaken this conclusion: we will come back
to this point later on.

Many attempts to simultaneously
explain both the responses have been carried out -- we restrict
our considerations to heavy and medium nuclei -- but the results
of these theoretical efforts are still largely unsatisfactory. Both
details and a critical discussion about them can be found, for
instance, in Ref.~\cite{BoGiPa-93-t}.

At lower $q$ calculations performed along the RPA or
Tamm-Dancoff approximation schemes improve the agreement
with the data
\cite{DeLeBr-85,BrDe-87,Ca84,DrCoWaSp-87,CoQuSmWa-88} as
far as collectivity can strongly influence the response. However, at
momenta of the order of 300 or 400 MeV/c this mechanism
cannot be effective anymore, but still the discrepancy between
the experimental data and the theory survives.

An interesting suggestion comes from the bag model: a partial
deconfinement mechanism could be effective inside nuclear
matter, leading to the  so-called ''swollen nucleon hypothesis'': the
corresponding change in the e.m. form factor could well account
for the quenching of the quasi elastic peak
(QEP)\cite{No-81,CeRoSh-84,CeRoSh-86}; the
experimental outcomes in the transverse channel remain,
however, unexplained: if the nucleon is swollen, $R_T$ should be
quenched too. A possible solution was suggested
by M. Ericson and Rosa-Clot\cite{ErRo-86}: in a mesonic model
the swollen nucleon can be seen as a bare nucleon surrounded by its
mesonic cloud, now feeling the presence of the other nucleons of
the medium so affecting in some unspecified way the photo-absorption.
It is easy to see that such a mechanism is channel
dependent, giving us a chance of explaining both the longitudinal and
transverse response functions in the frame of a mesonic theory.

We shall investigate this last topic in detail, examining in the
present
paper the longitudinal response function; we plan, in two other
following papers, to study the transverse response and the
nuclear sum rules as well.

Two crucial points need to be
clarified just from the beginning, in order to give a defined
content to the
words ``mesonic theory'': by one side a complete specification of
the dynamical model is needed -- this means to decide which
kinds of mesons and baryons are considered as true degrees of
freedom, to establish a lagrangian governing their dynamics
and, furthermore, to discuss which other quantities need to be
parametrized to obtain a realistic description of the system;
on the other side, a well defined theoretical scheme is asked for,
able to select which approximations can be safely carried out
to obtain a practically manageable and well behaved expansion
without violating general theorems.

Few years ago we proposed a scheme based on the Saddle Point Approximation
(SPA) applied to the generating functional for
a system of nucleons and pions\cite{AlCeMoSa-87}: this step can be performed
after the explicit integration of the fermionic degrees of freedom; in
the resulting expansion the ring-dressed meson propagators are the
mean field level of the theory, the higher order corrections being expressed
in terms of a loop expansion with respect to these ring-dressed boson
propagators.

In\cite{AlCeMoSa-87} we concentrated on a
relativistically covariant $\pi$-$N$ system, but the expansion there
derived was characterized by the topology of the diagrams: thus we are
free -- as we shall prove in the following -- to apply the same
scheme to the non relativistic restriction of the model and to
mesonic fields other than the pionic one.

In the present paper we shall include into the model
both the $\rho$-meson, displaying a remarkably different
behaviour in the
spin-transverse channel, and the $\Delta_{33}$ resonance, whose
contribution is relevant to a realistic description of the nuclear
dynamics. This will provide both the theoretical framework for a
forthcoming paper and a self-contained explanation of the
approach.

The results of a preliminary calculation were presented in
\cite{AlCeMoSa-90} in a purely
pionic framework. The theoretical expectation of strong
cancellations in the selected class of diagrams was supported by
the almost complete suppression of the one-loop corrections in the
isovector channel; however the longitudinal response was still overestimated.

In the present paper we extend the previous calculations to a
richer dynamical model by allowing the exchange of $\rho$ and
$\omega$ mesons and the excitation of $\Delta$-resonances in the
intermediate states. These contributions significantly improve the
agreement
with the experimental data.

The paper is organized as follows: in sect. \ref{sect2} the bosonic
loop expansion is presented and the one-loop corrections are
derived. The formalism is sufficiently general to apply to the
transverse channel as well, but the Meson Exchange Currents
(MEC) are not considered in the present paper: they will be
discussed when investigating the transverse (magnetic)
response. In sect. \ref{sect3} the dynamical model will be
presented and discussed. In sect. \ref{sect4} we shall illustrate the
details of the calculation and in sect. \ref{sect5} the results will be
presented. Finally in sect. \ref{sect6} we shall comment and discuss our
outcomes in connection with the experimental data. Some outlooks on the
perspectives of the present work are also presented.

\section{The Bosonic Loop Expansion}\label{sect2}

The functional approach to the nuclear Many-Body problem was
repeatedly described in previous
works\cite{AlCeMoSa-87,AlCeMo-89,Ce89-B,CeSa-89}, the
central idea being the projection of the action on a bosonic Hilbert
space via a functional integration of the fermionic degrees of
freedom. We studied in details in \cite{AlCeMoSa-87}
the e.m. response function within the
functional approach  for a system with nucleons and pions only.
Here we introduce two main modifications,
namely the non relativistic approximation and the inclusion of
those short range correlations (SRC) not coming from the exchange
of the mesons actually present in the model, to allow the use of the
presently available phenomenological models of low
momentum strong interactions, like,
e.g., the Bonn potential or whatever else mesonic theory.

To preserve coherence as far as possible, but still remaining
sufficiently general, we confine ourselves to a potential theory
with local interaction in the various particle-hole
channels with assigned spin, isospin and helicity.

The classical action (in terms of Grassmann variables) takes the
form
\begin{eqnarray}
A&=&\int dx\,dy\,\psi^\dagger(x)G_0^{-1}(x-
y)\psi(y)\label{lr1}\\
&&-{1\over2}\int
dx\,dy\,\sum_{i}\psi^\dagger(x)\Gamma_i\psi(x)
V_i(x-y)\psi^\dagger(y)\Gamma_i\psi(y)\nonumber
\end{eqnarray}
where the $\Gamma_i$ denote the relevant spin-isospin matrices
(to exemplify, in the spin-longitudinal  isovector channel
$\Gamma_i=-i\bbox{\sigma}\cdot\bbox\nabla \tau_i$) for which
we require a sort of orthogonality, namely
\begin{equation}
{\rm Tr}{\Gamma_i\Gamma_j}=n_i\delta_{ij}\,.
\label{lr2}
\end{equation}
We next couple the system to some external classical fields $J_i$
with the same quantum numbers of the corresponding
p-h channels and write the generating functional as
\begin{equation}
Z[J_i]=\int {\cal
D}[\psi^\dagger,\psi]e^{iA+i\sum_{i}\psi^\dagger\Gamma_i\psi
J_i}\,;
\label{lr3}
\end{equation}
the response function in a given channel follows from
\begin{equation}
\Pi^i(x,y)=-\left.\frac{\displaystyle \delta^2\log Z[J]}
{\displaystyle \delta J_i(x)\delta J_i(y)}
\right|_{J_i=0}
\label{lr4}
\end{equation}
and
\begin{equation}
R_i(q)=-\frac{1}{\pi}\Im \int d^4(x-y)e^{iq\cdot(x-y)}\Pi^i(x-y)\,.
\label{lr5}
\end{equation}
The thermodynamic limit is here understood, but it is
not in principle required: the present scheme, with
straightforward modifications, could apply to a finite system
as well; however the required computational effort makes
an actual calculation realistically not feasible.

The projection on a bosonic Hilbert space is carried out by means
of a Hubbard-Stratonovitch transformation
\cite{MoGaNaRa-74,Ke-70,Kl-78},
i.e., by exploiting the identity
\begin{eqnarray}
& e^{\displaystyle\,-\frac{\displaystyle i}{\displaystyle 2}
\int dx\,dy\,\psi^\dagger(x)\Gamma_i
\psi(x)V_i(x-y)\psi^\dagger(y)\Gamma_i\psi(y)}=\nonumber\\
& \sqrt{\det V_i}\int{\cal D}[\sigma_i]
e^{\displaystyle\,\frac{\displaystyle i}{\displaystyle 2}\int
dx\,dy\,\sigma_i(x)V_i^{-1}(x-y)\sigma_i(y)}\nonumber\\
&\times e^{\displaystyle \,i
\int dx\,\sigma_i(x)\psi^\dagger(x)\Gamma_i\psi(x)},\label{lr6}
\end{eqnarray}
then substituting (\ref{lr6}) into (\ref{lr3}) and carrying out
the remaining gaussian integration over the
fermionic fields. We end up
(after a shift in the integration variable $\sigma_i$) with
\begin{equation}
Z[J_i]=\exp\left\{\frac{i}{2}\sum_{i}J_iV^{-1}_iJ_i\right\}
\int{\cal D}[\sigma]\exp\left\{iA^B_{\rm eff}\right\}
\label{lr7}
\end{equation}
where
\begin{eqnarray}
A^B_{\rm eff}&=&{1\over 2}\sum_{i}\sigma_i
V^{-1}_i\sigma_i-i{\rm Tr}\sum_{n=1}^{\infty}{1\over
n}\left[\sum_{i}\sigma_i\Gamma_iG_0\right]^n
\nonumber \\ && \qquad
-\sum_{i}\sigma_iV_i^{-1}J_i\;.
\label{lr8}
\end{eqnarray}
This effective action has been derived in the frame of a potential
theory. However, a mesonic model leads to the same form of the
generating functional with the potentials $V_i$ replaced by
meson propagators; the fields $\sigma_i$ can then be interpreted
as the true mesonic fields. Thus, thanks to this unifying aspect
of the formalism, we can proceed without distinguishing between
mesonic and potential theories and attribute to the $V_i$'s the
appropriate meaning, potential or meson propagator, directly
when needed. In eqs. (\ref{lr7}) and (\ref{lr8}) the quantity
$V_i^{-1}$ appears, so raising the delicate question of its
existence: this problem can be formally solved
by means of the change of variable $\sigma_i\to V_i\sigma_i$ in
the functional integrals, but we prefer to keep the present
form to preserve to $<\sigma_i>$ the meaning of expectation value
of the underlying mesonic fields.

The next step is the evaluation of the last path integral in
eq.~(\ref{lr7}). We adopt the semiclassical expansion, i.e.,
the path integral is evaluated within the Stationary Phase
Approximation -- or Saddle Point Approximation in the Euclidean
word, in both cases SPA. This amounts to impose the stationarity
of the effective action $A^B_{\rm eff}$ with respect to arbitrary
variations of the fields $\sigma_i$:
\begin{eqnarray}
&  J_j(x) = \sigma_j(x)\label{lr9} \\
& -{\rm Tr}\int dy\,V_j(x-y)\sum_{n=1}^{\infty}\int dy_1\dots
dy_nG_0(y-y_1)
\nonumber\\ &\cdot
\sum_{i}\Gamma_i\sigma_i(y_1)
G_0(y_1-y_2)\dots\sum_{k}\Gamma_k\sigma_k(y_n)
\nonumber\\ &\times
G_0(y_n-y)\Gamma_j\;.
\nonumber
\end{eqnarray}
The solutions of eq. (\ref{lr9}) are clearly functionals of
the external sources $J_j$; they will be denoted from now on by
$\sigma^{\rm RPA}_j$ because of the link, that will become clear
in the following, with the Random Phase Approximation -- or
more precisely, with the ring approximation, as it is sometime
referred to.

A peculiar role is played by the
scalar-isoscalar field (if any), because it brings into the theory the
only possible tadpole; if so, a spontaneous symmetry
breaking occurs and a shift in the corresponding integration
variable is needed to force back the field expectation value to
zero; this procedure entails a finite renormalization of the
non interacting nucleon Green's function $G_0$ with the Hartree
potential generated by the tadpole.
In the present work we do not introduce the $\sigma$-meson:
then this procedure is not needed.

We can solve the SPA equation in the form of Volterra expansions for
$\sigma^{\rm RPA}_j$ as functional of
$J_j$. Having ruled out the term with $n=1$ in eq.~(\ref{lr9}), the
$0^{th}$-order term of the expansion vanishes and we can write
\footnote{We use, when no confusion can arise, the convention that  a sum
is understood when some spin-isospin index is repeated {\bf two
and only two} times. In the same way a repeated space-time
variable is meant to be integrated.}:
\begin{eqnarray}
&\sigma^{\rm RPA}_i(x)=\label{lr10}
\sum_k\int dy\,A^i_k(x|y)J_k(y)+
\\ &+{1\over 2}\sum_{km}\int dy dz \,
B^i_{km}(x|y,z)J_k(y)J_m(z)+{\cal O}(J^3) \nonumber
\\ &\equiv
A^i_k(x|y)J_k(y)+{1\over 2}B^i_{km}(x|y,z)J_k(y)J_m(z)
\nonumber \\ & \qquad\quad
+{\cal O}(J^3) \nonumber\;.
\end{eqnarray}
Substituting the form (\ref{lr10}) into (\ref{lr9}) and collecting
the terms linear in $J_i$ one gets for $A^i_k$ the equation
\begin{eqnarray}
A^i_k(x|y)&=&n_i\int
dz\,dt\,V_i(y-z)\Pi^{(2)}(z-t)A^i_k(t|y)
\nonumber\\ &&\qquad
+\delta_{ik}\delta(x-y)
\label{lr11}
\end{eqnarray}
$\Pi^{(2)}$ denoting the Lindhard function.
The reasons of this unusual notation will become clear later, see
eq (\ref{lr20}). Then, since the RPA-dressed potential (or meson
propagator) reads
\begin{equation}
\Delta_{\rm RPA}^{i}={1\over V_i^{-1}-n_i\Pi^{(2)}}
\label{lr12}
\end{equation}
the solution for $A^i_k$ takes the form
\begin{equation}
A^i_k(x|y)=\Delta_{\rm RPA}^{i}(x-z)V_i^{-1}(z-
y)\delta_{ik}\;.
\label{lr13}
\end{equation}
It is remarkable that formally
\begin{equation}
A={1\over 1-nV\Pi^{(2)}}
\label{lr14}
\end{equation}
so that if no interaction in the $j^{th}$ channel is allowed
($V_j=0$), then
\begin{equation}
A^j_k(x|y)=\delta_{jk}\delta(x-y)\;.
\label{lr15}
\end{equation}

The equation for $B$ reads
\begin{eqnarray}
&B^i_{jk}(x|y,z)=n_i V_i(x-u)
\Pi^{(2)}(u-v)B^i_{jk}(v|y,z)+\label{lr16}\\
&2{\rm Tr} \Gamma_i\Gamma_j\Gamma_k V_i(x-u) G_0(u-v)
\left(\Delta_{\rm RPA}^{j}V^{-1}_j\right)(v-y)
\nonumber\\
&\cdot G_0(v-t)\left(\Delta_{\rm RPA}^{k}V^{-1}_k\right)(t-z)
G_0(t-u)\nonumber
\end{eqnarray}
with the immediate solution
\begin{eqnarray}
B^i_{jk}(x|y,z)&=&2{\rm Tr}\,\Gamma_i\Gamma_j\Gamma_k
\Delta_{\rm RPA}^{i}(x-u)G_0(u-v)
\label{lr17}\\ &&
\quad\cdot\left(\Delta_{\rm RPA}^{j}V^{-1}_j\right)(v-y)G_0(v-t)
\nonumber\\ &&\qquad\cdot
\left(\Delta_{\rm RPA}^{k}V^{-1}_k\right)
(t-z)G_0(t-u)\nonumber\;.
\end{eqnarray}

Having solved the SPA equations, the generating functional reads
\begin{eqnarray}
Z[J_j]&=&e^{\displaystyle\frac{i}{2}\sum_i J_iV^{-1}_iJ_i}
e^{\displaystyle\, iA^B_{\rm eff}[\sigma^{\rm RPA}_j]}
\nonumber\\&&\quad\times\det
\left\{\frac{\delta^2 A^B_{\rm eff}[\sigma]}
{\delta\sigma_i(x)\delta\sigma_j(y)}
\Biggm|_{\sigma_k=\sigma^{\rm RPA}_k}\right\}^{-{1\over2}}
\label{lr18}
\end{eqnarray}
where the exponent describes the mean field propagation while
the determinant is responsible for the one-loop (quadratic)
corrections [both discrete and continuous indices must be
considered in the evaluation of the symbolic determinant in
Eq.~(\ref{lr18})].

Let us consider first of all the mean field response functions: to
obtain the linear response it suffices to replace into $A^B_{\rm
eff}$ the SPA solutions for $\sigma_i$ up to the first order in $J_i$;
only the term with $n=2$ of the sum in (\ref{lr8}) comes into play
and then
\begin{equation}
-\left.\frac{\delta^2\log Z}{\delta J_i\delta J_i}\right|_{J_i=0}
=\frac{\displaystyle \Pi^{(2)}}{\displaystyle 1-n_iV_i\Pi^{(2)}}
\label{lr19}
\end{equation}
i.e., the response function is given by its RPA approximation.

When the potential is vanishing the mean field response
reduces to the FFG one.
To exemplify, in the case we are interested in -- the response to a
scalar-isoscalar plus scalar-isovector probe, a charge
longitudinal $\gamma$ -- and if the effective
interaction carries only the $\pi$, $\rho$ or $\omega$ quantum
numbers both the mean fields turn out  to coincide with the FFG
ones. Instead, in the spin transverse channel (the magnetic one)
the $\rho$-meson propagation is able to dress at the mean field
level the FFG response.

Let us now come to the one-loop corrections:
the fermionic loops up to $n=4$ are relevant. We define,
following the notations of \cite{CeSa-88,CeCoCoSa-93-t}, the
functions
\begin{eqnarray}
\Pi^{(2)}(x,y)&=&G_0(x-y)G_0(y-x)\label{lr20}\\
\Pi^{(3)}(x,y,z)&=&G_0(x-y)G_0(y-z)G_0(z-x)\label{lr21}\\
\Pi^{(4)}(x,y,z,t)&=&G_0(x-y)G_0(y-z)G_0(z-t)
\nonumber\\ && \qquad\times
G_0(t-x)\label{lr22}\;,
\end{eqnarray}
which are symmetric for cyclic permutations of the
arguments: the second derivatives of $A^B_{\rm eff}$ result:
\begin{eqnarray}
&\frac{\displaystyle\delta^2A^B_{\rm eff}}
{\displaystyle\delta\sigma_i(x)\delta\sigma_j(y)}=
V_i^{-1}(x-y)\delta_{ij}-
\nonumber
\\
&\sigma_k(z)\left\{\Gamma_i\Gamma_j\Gamma_k
\Pi^{(3)}(x,y,z)+
\Gamma_j\Gamma_i\Gamma_k\Pi^{(3)}(y,x,z)\right\}\nonumber
\\
&-\sigma_k(z)\sigma_m(t)
\left\{\Gamma_i\Gamma_j\Gamma_k\Gamma_m\Pi^{(4)}(x,y,z,t)+
\right.\nonumber\\
&\left.\Gamma_j\Gamma_i\Gamma_k\Gamma_m\Pi^{(4)}(y,x,z,t)+
\Gamma_i\Gamma_k\Gamma_j\Gamma_m
\Pi^{(4)}(x,z,y,t)\right\}\;.\nonumber
\end{eqnarray}
Next using the well known property
$$\det X=e^{{\rm Tr}\log X}$$
the one loop contribution to the polarization propagator becomes
\begin{eqnarray}
\Pi^{\rm 1-loop}_i(x,y)&=&-{i\over 2}
\frac{\displaystyle\delta^2}{\displaystyle\delta J_i(x)\delta
J_i(y)}
\nonumber
\\ \nonumber &&
\quad\cdot
\left[{\rm Tr}\log\frac{\displaystyle\delta^2
A^B_{\rm eff}}{\displaystyle\delta\sigma_i(x)\delta\sigma_j(y)}
\Biggm|_{\sigma_i=\sigma^{\rm RPA}_i}\right]_{J_k=0}\,.
\end{eqnarray}
The logarithm is required up to the order $J^2$: thus we
multiply in the exponent by $V_i$ -- a quantity not
affecting the response because independent from $J$ -- and we
expand the logarithm up to the second order before deriving it
with respect to the external sources. A tedious but
straightforward calculation provides
\begin{eqnarray}
&\Pi^{\rm 1-loop}_i(x,y)=
\nonumber
i\left(\Delta_{\rm RPA}^{i}V_i^{-1}\right)(x-z)
{\rm Tr}\Gamma_i\Gamma_k\Gamma_i\Gamma_k
\\ &
\Pi^{(4)}(z,u,t,v)V_k(u-v)\left(\Delta_{\rm RPA}^{i}V_i^{-
1}\right)(t-y)\nonumber\\ &
+i\left(\Delta_{\rm RPA}^{i}V_i^{-1}\right)(x-z)V_k(u-v)
{\rm Tr}\Gamma_i\Gamma_k\Gamma_k\Gamma_i\nonumber\\
&
\left\{\Pi^{(4)}(z,u,v,t)+\Pi^{(4)}(t,u,v,z)\right\}\left(\Delta_{\rm
RPA}^{i}V_i^{-1}\right)
(t-y)\nonumber\\ &
+i\left(\Delta_{\rm RPA}^{i}V_i^{-1}\right)(x-z)
{\rm Tr}\Gamma_i\Gamma_j\Gamma_k\Pi^{(3)}(z,t,u)V_j(t-
v)\nonumber\\ &
V_k(u-w)\left\{{\rm
Tr}\Gamma_i\Gamma_k\Gamma_j\Gamma_j\Pi^{(3)}(s,w,v)
\right.\nonumber\\ &
\left.+{\rm
Tr}\Gamma_i\Gamma_j\Gamma_k\Pi^{(3)}(s,v,w)\right\}
\left(\Delta_{\rm RPA}^{i}V_i^{-1}\right)(s-y)\nonumber\;.
\end{eqnarray}
The corresponding Feynman graphs are plotted in fig. \ref{fig1}.

The theory, in the form here presented, has been developed for
a system of nucleons interacting either through a potential or
through a meson exchange, without explicitly including the
possible excitation of baryonic resonances. This
drawback is overcame by replacing the fermionic
fields with fermionic multiplets, whose components are the true
field operators for the various kinds of baryons. The diagrams so
generated are topologically identical to those of fig. \ref{fig1}, but
each fermionic line stands now either for a nucleon or for a
$\Delta$, or for whatever else resonance we choose to include into
the dynamical model.

\section{The Dynamical Model}\label{sect3}

The approximations here described are sufficiently general
to apply to various dynamical schemes. We now describe the
one we are going to follow.

The complete determination of the model requires the
specification of two ingredients -- the form of
the current and the potential we choose.

Concerning the first point, we recall that the response to an e.m.
probe in the longitudinal channel is usually described, in a
non relativistic context, by its nucleonic part, neglecting any
contribution coming from the MEC [contact term, pion-in-flight,
direct $\Delta$-excitation]. This assumption seems
quite acceptable, and we shall follow it in this paper: in principle,
however, we cannot exclude a significant contribution coming
from the MEC, even in the longitudinal channel, when relativity is
correctly accounted for.
In the one-loop approximation, in fact, we expect large
cancellations between the various contributions -- and
they were found indeed in the isovector channel in a preliminary
calculation\cite{AlCeMoSa-90}: in such a
case even small contributions could become relevant. Thus the
corrections coming from the MEC in the longitudinal channel are
for the moment neglected, but they will be object of a future
investigation.

To study the longitudinal response we only need two external
sources, namely $\Gamma_{T=0}=I$ and $\Gamma_{T=1}=\tau_3$,
the full response being at the FFG level, with obvious notations,
\begin{equation}
R_L={1\over 4}R_{T=0}+{1\over 4}R_{T=1}\;.
\label{lr26}
\end{equation}

Both the response functions in (\ref{lr26}) contain, of course, the usual
Sachs form factor $G_{E\,p}$.

Next we come to the second, and more delicate,
problem, namely the choice of the $V_i$. Three mesonic fields are known to
dominate the nuclear dynamics -- as it can be argued for instance
from the $N$-$N$ phase shifts analysis leading to the Bonn
potential\cite{Ho-81,MaHoEl-87} -- namely those of
$\pi$, $\rho$ and $\omega$ mesons. To follow this scheme
we shall only consider those $V_i$ pertaining to the isovector
spin-longitudinal and spin-transverse channels and to the isoscalar
spin-transverse one. This does not mean that we completely
neglect the effective interaction in other channels,
which should be thought as arising, largely, from the two
mesons exchange, a process accounted for in our scheme by the
diagrams d) and e) of fig. \ref{fig1}. It is important to
remember, for instance, that the $\sigma$ meson -- carrying the
scalar-isoscalar part of the interaction -- was introduced in
the past to simulate the exchange of two pions with the
simultaneous excitation of one (or two) of the intermediate
nucleons to a $\Delta$. In the present scheme these contributions
-- often referred to as the ``box diagrams'' in the language of the
Bonn potential -- are explicitly accounted for.

We now examine the model interaction in the various channels,
starting from the isovector spin-longitudinal one. There, since the
pioneering works of Migdal on pion
condensation\cite{Mi-71,Mi-78}, the effective interaction is
described by the one pion exchange plus a short range
contribution schematized by the Landau parameter $g^\prime$
which simulates ``whatever else can happen'' in that
channel\cite{DiFaMeMu81a,DiFaMeMu81b,CeSa90}. In the present
scheme this assumption is too poor for at
least two reasons: the first is that in an one loop calculation a
constant potential leads to (obviously
spurious) divergences, and the same happens for a p--wave static
pion exchange potential (see Ref.~\cite{Ce-93-t} for more
details); the usual $\pi NN$ vertex form factor obviously removes
the divergence, but it leaves unaltered the unphysical high
momentum behaviour of the interaction, which in this way is
simply hidden. The second reason is that, again at the one loop
level, some diagrams can occur with two consecutive Landau
effective interactions attached to the same fermionic line [this is
immediately seen to
happen, again, for diagrams d) and e) of fig.~\ref{fig1}, but the
same
is also true for
diagrams b) and c)]: this means that some double counting can, in
principle, be present in the calculation; thus we must spend some
more words to better specify the physical meaning we attribute to
$g^\prime$ at the 1--loop level.

When the momentum carried by one meson [or potential] line is
not limited by the kinematics, as instead it
happens at the mean field [0-loop] level, it is necessary to
account for a major effect, usually referred to as  ``short
range correlations'' [SRC]\cite{BrBaOsWe-77}: they prevent, for instance,
the pion to be exchanged between nucleons too near the one to
the other. This effect can be ascribed, within a mesonic theory,
to a short range repulsive interaction, thought as a heavier meson
exchange occurring before and/or after the pion exchange. In
fig.~\ref{fig2} the exchange of a pion
between correlated nucleons is shown, the wiggly line denoting a
$G$-matrix obtained only from the repulsive short range interaction.
Remarkably such an effect could be well described by means of
seemingly completely different models, like, e.g.,
exotic quark configurations; within the present scheme it is
simpler to identify and understand the microscopical origin of the channel
dependence of the effect we are looking for.

If ${\cal V}_\pi(r)$ denotes the one pion
exchange potential in configuration space, the four diagrams of fig.
\ref{fig2} are well simulated by $g(r){\cal
V}_\pi(r)$, $g(r)$ being the pair correlation function [more
precisely we should evaluate the matrix element of ${\cal V}_\pi$
with correlated wave functions, i.e., solutions of the Bethe-Goldstone
equation, instead making use of plane
waves\cite{CeDi-81}: the previous parametrization corresponds to
the static limit].

Within this scheme an overcounting of diagrams arises
when two successive correlated pions are exchanged (as it
happens in all but the first diagrams of fig. \ref{fig1} when
the first [diagrams d) and e)] or second [diagrams b) and c)]
term only of the RPA series is considered). The diagrams
presenting overcounting are displayed in fig.
\ref{fig3}.

In the static limit, however, two successive meson
exchanges are $\propto g(r)^2{\cal V}^2_\pi(r)$, while
dropping the double counting roughly corresponds to
$\propto g(r)^{3\over2}{\cal V}^2_\pi(r)$. Thus, due to the form
of $g(r)$, the effect of the over counting is to amplify the
suppression of the short range part of $V_\pi(r)$, i.e., in Fourier
transforms, of its high momentum components. This outcome is
unavoidable from a formal point of view: it originates from
handling an effective interaction as a true potential.

Brown et al. suggested for $g(r)$ the very simple
form
\begin{equation}
g(r)=1-j_0(q_cr)
\label{lr27}
\end{equation}
(with $q_c\sim m_\omega$)
which leads to the following expression of the correlated potential
in Fourier transforms:
\begin{equation}
g(r){\cal V}_\pi(r)\Longrightarrow{\cal V}_\pi({\bf q})-
{\int \frac{d^3 k}{(2\pi)^3}}
{2\pi^2\over q_c^2}\delta(|{\bf k}|-q_c){\cal V}_\pi({\bf
k}-{\bf q}),
\label{lr28}
\end{equation}
widely employed in the literature\cite{OsSa-87} for
two main reasons: it gives a good description of the elastic
pion-nucleus
scattering\cite{OsWe-79,OsWe-79a} and it corresponds, in
the Landau limit and accounting also for the correlated $\rho$
exchange, to a value for $g^\prime\sim 0.65$, consistent
with many model-independent analysis.
Furthermore, it cuts down the high momentum components of the
correlated potential because of the cancellation between the two
terms in (\ref{lr28}) when $|{\bf q}|\gg q_c$.

We shall adopt in the following a different form -- maintaining
however the same philosophy --  by writing
the potential in the isovector spin-longitudinal channel as
\begin{equation}
V_\pi(q)={f^2_{\pi NN}\over m_\pi^2}\left\{g^\prime_L({\bf
q})-
{{\bf q}^2\over m^2_\pi+{\bf q}^2}\right\}v_\pi^2(q^2)\,;
\label{lr29}
\end{equation}
and in the isovector spin-transverse as
\begin{equation}
V_\rho(q)={f^2_{\pi NN}\over
m_\pi^2}\left\{g^\prime_T({\bf q})-
C_\rho{{\bf q}^2\over m^2_\rho+{\bf q}^2}\right\}v_\rho^2(q^2)\;.
\label{lr30}
\end{equation}
A momentum dependence has been attributed to
$g^\prime_L$ and
$g^\prime_T$, such that
\begin{equation}
g^\prime_L({\bf q})\buildrel {q\to\infty}\over\longrightarrow
1\qquad\qquad g^\prime_T({\bf q})\buildrel
{q\to\infty}\over\longrightarrow C_\rho\;.
\label{lr31}
\end{equation}
Actually the functional forms we have chosen are
\begin{eqnarray}
g^\prime_L(q)&=&1+(g_0^\prime-1)\left[{q_{c\,L}^2\over
q_{c\,L}^2+q^2}\right]^2
\label{lr32}\\
g^\prime_T(q)&=&C_\rho+(g_0^\prime-
C_\rho)\left[{q_{c\,T}^2\over
q_{c\,T}^2+q^2}\right]^2\label{lr33}\;.
\end{eqnarray}
With these expressions we obtain the required
high momenta cancellations and we force the same Landau limit
$g_0^\prime$ in both channels, as it should be.
Following \cite{OsSa-87,OsWe-79,OsWe-79a}
the value of $q_{c\,L}$ should be chosen around $m_\omega$, but,
because of the previous considerations about the double counting
problem we suggest as an acceptable compromise a slight increase
of the value of $q_{c\, L}$ to
$q_{c\,L}\sim 0.8\div 0.9~{\rm GeV/c}$: in this way we effectively
remove the over counting problem, without altering the low-$q$
behaviour of the effective interaction. We have no
phenomenological indications about the precise value $q_{c\,T}$
should have: surely it should lie between $\sim 1~{\rm GeV/c}$ and
$\sim 2~{\rm GeV/c}$. The lower limit comes when considering that
obviously the effect of the SRC for the $\rho$-meson
exchange should be sensible at shorter distances with respect to
$1/m_\rho$; the upper one is roughly determined by the
inverse size of the $NN\rho$-vertex. Apart from these wide limits
we assume $q_{c\,T}$ essentially as a free parameter.

Finally we come to the isoscalar spin-transverse channel. As we
are going to discuss in the following, phenomenology does not
impose severe constraints on the
$\omega$-meson propagation in nuclear matter. Thus we simply
assume
\begin{equation}
V_\omega(q)=-{f^2_{\pi NN}\over m_\pi^2}C_\omega
{{\bf q}^2\over m^2_\rho+{\bf q}^2}v_\omega^2(q)\,.
\label{lr34}
\end{equation}

The form factors in (\ref{lr29}), (\ref{lr30}) and (\ref{lr34})
are in the usual dipole
form
\begin{equation}
v(q^2)={\Lambda_m^2-m_m^2\over \Lambda_m^2 + {\bf q^2}
-q_0^2}
\label{lr34.1}
\end{equation}
with cut-offs, respectively, chosen to be $\Lambda_\pi =
1.3~{\rm GeV/c}$ $\Lambda_\rho = 2.5~{\rm GeV/c}$ and
$\Lambda_\omega=1.5~{\rm GeV/c}$.

Finally let us discuss the coupling constants. We put as usual
$f^2_{\pi NN}/4\pi=0.08$ and $f^2_{\pi N\Delta}/4\pi=0.32$;
the value we use for $f_{\pi N\Delta}$ is commonly employed in
the literature, somehow higher than the value obtained by
the $N$-$N$ phase shift analysis carried out in constructing the
Bonn potential, but lower than the results of a recent
fit\cite{CeChDi-87}. It is remarkable that the precise value of this
parameter depends on the details of the underlying dynamical
model: within the Bonn potential approach, in our opinion, is
necessary to underestimate the $\pi N \Delta$ coupling, to
effectively account for the $\Delta$-$\Delta$ repulsion inside a
box diagram\cite{CeCoLo-89,CeMaSa-91}. Concerning
the best fit of Ref.~\cite{CeChDi-87}, the higher value of the
coupling constant was accompanied by a reduction of the cut-off
so that the net effect is roughly the same.

Coming to vector mesons, we write the relativistic interaction
lagrangian in the form
\begin{equation}
g_V\overline{\psi}\gamma_\mu\psi\phi_V^\mu+{f_V\over
4m}\overline{\psi}\sigma_{\mu\nu}\psi\left(\partial^\mu\phi_V^
\nu-
\partial^\nu\phi_V^\mu\right)
\label{lr35}
\end{equation}
which reduces, in the non relativistic limit, to
\begin{equation}
i{(f_V+g_V)^2\over 4m}\psi^\dagger\bbox{\sigma}\times {\bf
q}\psi\cdot
\bbox{\phi}_v\,.
\label{lr36}
\end{equation}
Let us finally define $C_V$ as
\begin{equation}
{(f_V+g_V)^2\over 4m}=C_V {f^2_{\pi NN}\over m_\pi^2}
\label{lr37}
\end{equation}
so to give a meaning to eqs. (\ref{lr30}) and (\ref{lr34}).

Using the values given by Ref.~\cite{HoPi-75} one gets
$C_\rho\sim 2.3$.
The coupling constant of the $\omega$ is not well known. H\"oler
et al. \cite{Ho76}  values correspond to
$0.83<C_\omega<2.5$; Grein \cite{Gr-77} and Grein and Kroll
\cite{GrKr-80} provide respectively $C_\omega = 0.83$ and 0.56.
Finally the $N$-$N$ phase shift analysis of Ref.~\cite{MaHoEl-87}
gives $C_\omega=1.5$, the value we have adopted in the present
work.
As it will become clear from our results the uncertainty over this
parameter does not propagate too much on the final results,
being the overall $\omega$--contribution
small. Also the possible presence of a residual SRC interaction in
this channel -- that we {\bf do not} include in the present
calculation -- should not influence the final results.

Finally, the $\rho N\Delta$ coupling is also largely unknown. The
analysis of Ref.~\cite{MaHoEl-87} provides $C_\rho^\Delta=2.1$
but other values are also compatible.
Here we have chosen $C_\rho=2.3$ both for $N$ and $\Delta$.
The corresponding cut-offs are also chosen according to the Bonn
potential parameters.

We remark that in the present approach the practically unknown
$\Delta\Delta$-meson vertices are not required.

\section{Evaluation of the diagrams}\label{sect4}

The diagrams corresponding to the one loop corrections, described
in Sect.~\ref{sect2}, are those of Fig.~\ref{fig1} (without $\Delta$
lines), or those with one $\Delta$ line in Fig.~\ref{fig4cd} and with
two $\Delta$ lines in Fig.~\ref{fig4e}. Everywhere the wiggly line
denotes a RPA-dressed meson ($\pi$, $\rho$ or $\omega$) and a
double solid line stands for a $\Delta$ propagator. Each term is a
Feynman diagram, i.e., we could further reduce it in Goldstone
diagrams by distinguishing all the possible time-orderings. Such a
procedure entails an enormous increase in the number of
terms required: for instance, diagrams a), b) and c) generate 24
Goldstone diagrams each, while 720 Goldstone diagrams come
from d) and e) [remember that the internal mesonic lines are RPA
dressed, so they are not forced to connect equal-times couples of
points]. Since, furthermore, different particles may correspond to
the internal mesonic lines,
diagrams a), b) and c) must be evaluated for three different cases
while d) and e) for nine: the explicit evaluation of all the
separated Goldstone diagrams is a hopeless task.

We can, however, bypass this problem without loosing any
information by means of the results of
Refs.~\cite{CeSa-88,CeCoCoSa-93-t}, where we caried out
the analytic evaluation in momentum space
of the functions $\Pi^{(3)}$ and $\Pi^{(4)}$ describing a
fermionic loop with 3 or 4 external legs. We proved that only one
$\theta$-function is truly necessary when dealing with the generic {\em
Feynman} diagram $\Pi^{(n)}$: the information carried by
the remaining $\theta$'s can be transferred to the analytical
properties of the diagram in momentum
space. Remarkably, this result can be proved by
individually altering each Goldstone diagram, in such a way that
their sum remains unaltered. Thus the information
about the the single Goldstone diagram contributions are lost, but
the overall Feynman diagram can be {\bf exactly} evaluated.
Furthermore it is possible to choose a
particular energy-momentum region where the analytical
properties of the diagram are simply defined (this is typically a
high energy region): there the loop integral can be explicitly
carried out and then one is able to come back to the other
kinematical regions through the study of the analytical properties
of the whole diagram.

In momentum space we shall employ for the fermionic loops we
are interested in the notation $\Pi^{(3)}(p,q)$ and
$\Pi^{(4)}(p,q,k)$, $p$, $q$, $k$ being the momenta
entering the diagram, in clockwise order [the last momentum
entering each diagram is obviously fixed by momentum
conservation].

The function $\Pi^{(3)}(q_1,q_2)$, entering the diagrams d) and e),
can be expressed as
\begin{eqnarray}
&&\Pi^{(3)}(q_1,q_2)=m^2\left[I^3(q_1,q_2)+I^3(-q_1,q_2-
q_1)\right.\nonumber\\
&& \qquad\qquad\qquad \left.+I^3(-q_2,q_1-
q_2)\right]\label{lr37.1}
\end{eqnarray}
where the function $I^3$ is in turn defined as
\begin{eqnarray}
I^3(q_1,q_2)&=&\left.\frac{1}{(2\pi)^3q_1q_2\sin^2\chi}
\right\{
(y_1\cos\chi-y_2)\cdot\label{lr37.2}\\
&\cdot&\log\frac{y_1-k_F}{y_1+k}+(y_2\cos\chi-y_1)
\log\frac{y_2-k_F}{y_2+k}\nonumber\\
&-&\left.\sqrt{\Delta}\log\frac{y_1y_2-k_F^2\cos\chi
+k_F\sqrt{\Delta}}{y_1y_2-k_F^2\cos\chi
-k_F\sqrt{\Delta}}\right\}\nonumber\,.
\end{eqnarray}
In the last equation $q_{1,2}$ denotes the norm of the
3-vector $\bf q_{1,2}$ and $\chi$ is the angle between them.
Furthermore
\begin{equation}
y_{1,2}=\frac{mq^0_{1,2}}{q_{1,2}}-\frac{q_{1,2}}{2}
\label{lr37.3}
\end{equation}
and
\begin{eqnarray}
G^2 &=& y_1^2-2y_1y_2\cos\chi+y_2^2\\
\Delta &=& G^2-k_F^2\sin^2\chi\;.
\label{lr37.4}
\end{eqnarray}
The diagrams a), b) and c) require two particular cases of
$\Pi^{(4)}$,
which can be expressed in terms of $\Pi^{(3)}$, namely
\begin{eqnarray}
\Pi^{(4)}(k,q,-k)&=&-\frac{m^3}{{\bf k}\cdot{\bf q}}\left\{
\Pi^{(3)}(k,q)+\Pi^{(3)}(-k,-q)\right.\label{lr37.5}\\
&&-\left.\Pi^{(3)}(k,-q)-\Pi^{(3)}(-k,q)\right\}
\nonumber\\
\Pi^{(4)}(k,q,-q)&=&\left(\frac{\partial~}{\partial
k^0}+\frac{\partial
{}~}{\partial q^0}\right)\Pi^{(3)}(k,q)\;.\label{lr37.6}
\end{eqnarray}
The reader is referred to \cite{CeSa-88,CeCoCoSa-93-t} for further
details on the analytical extension of the formulae.

The diagrams f) to k) can be evaluated analytically too, if the
approximation
\begin{equation}
{m_\Delta m\over m_\Delta-m}\gg {k_F^2\over 2m}
\end{equation}
is assumed to hold: needless to say at the usual saturation density
this is near to be exact. The corresponding
formulae are not given here for obvious reasons of brevity. The
reader is again referred to \cite{CeCoCoSa-93-t}, where all the
details are given: the analytical functions required in this case are
as simple as the former ones.

The relevant point is that all the fermionic loop integrations have
been now analytically performed, resulting in easily manageable
functions. Thus the calculation of diagrams a) to k) reduces to a
4-dimensional integral over a known analytical function (actually
the integral is 3-dimensional because the integration over the
azimuthal angle is trivial) and the sum of the many thousands of
Goldstone diagrams required by our theoretical scheme is
translated to the sum of 51 Feynman diagrams each one
expressed as a 3-dimensional integral [the possible
exchange of $\pi$, $\rho$ and $\omega$ being accounted for].

Explicitly diagram a) reads
\begin{equation}
\Pi^{(a)}(k)=i{\rm
Tr}\Gamma_{em}\Gamma_i\Gamma_{em}\Gamma_i
{\int \frac{d^4 q}{(2\pi)^4}}\Pi^{(4)}
(k,q,-k)V_i(q)
\label{lr38}
\end{equation}
with $\Gamma_{em}=(1+\tau_3)/2$; $\Gamma_i$ and $V_i$
are the vertex and the potential pertaining to the exchanged
mesons; diagrams b) and c) read respectively
\begin{equation}
\Pi^{(b)}(k)=i{\rm
Tr}\Gamma_{em}\Gamma_i\Gamma_i\Gamma_{em}
{\int \frac{d^4 q}{(2\pi)^4}}\Pi^{(4)}
(k,q,-q)V_i(q)
\label{lr39}
\end{equation}
and
\begin{equation}
\Pi^{(c)}(k)=\Pi^{(b)}(-k)
\label{lr40}
\end{equation}
Finally diagrams d) and e) are given by
\begin{eqnarray}
\Pi^{(d)}(k)={\int \frac{d^4 q}{(2\pi)^4}}{\rm
Tr}\Gamma_i\Gamma_{em}\Gamma_j\Pi^{(3)}(q,k)V_i(q)
\nonumber \\
V_i(q+k){\rm Tr}\Gamma_j\Gamma_{em}\Gamma_i\Pi^{(3)}(-k,-
q)
\label{lr41}
\end{eqnarray}
and
\begin{eqnarray}
\Pi^{(e)}(k)={\int \frac{d^4q}{(2\pi)^4}}{\rm
Tr}\Gamma_i\Gamma_{em}\Gamma_j\Pi^{(3)}(q,k)V_i(q)
\nonumber\\
V_i(q+k){\rm Tr}\Gamma_i\Gamma_{em}\Gamma_j\Pi^{(3)}(-q,-
k)
\label{lr42}
\end{eqnarray}
where we explicitly indicated that two different particles ($i$ and
$j$) can be exchanged. Diagrams f) to k) have the same structure
of eqs.~(\ref{lr41}) and (\ref{lr42}),but the proper expressions
for the various $\Pi^{(3)}$ functions must be
used\cite{CeCoCoSa-93-t}, according to which line in the diagram
pertains to a nucleon or to a $\Delta$.

The last step before the numerical evaluation of the integrals is
the calculation of the spin--isospin traces. In the
channels we are considering the various $\Gamma$'s are respectively
$\bbox{\sigma}\cdot {\bf
q}\tau_\mu$,
$(\bbox{\sigma}\times {\bf q})_i\tau_\mu$ and
$(\bbox{\sigma}\times {\bf q})_i$ for $\pi$, $\rho$ and
$\omega$-emission respectively. For the case of a $N\Delta$
transition vertex, the matrices $\bbox\sigma$ and $\tau_\mu$
will be replaced by the proper spin-isospin transition ones,
usually denoted by ${\bf S}$ and $T_\mu$.

The traces reduce to a coefficient in front of the diagram,
including, if case, an angular dependence.
Those pertaining the diagrams a), b) and c) are given in table
\ref{tab1}, where the contributions of the isoscalar and of the
isovector part have been separated. Diagrams d) to k) require the
study of 72 different cases, because in each of them we must
distinguish  the particles exchanged inside (9 possibilities). We can
simplify the situation observing that the diagrams can be
characterized as follows:
\begin{enumerate}
\item the directions of the momentum flow in the two loops of
diagrams d), f), h), j) are opposite (we shall call these
``direct correlation diagrams''), while for e), g), i), k) case they are
the same
(``exchange correlation diagrams'');
\item the diagrams are characterized by the exchanged particles,
$\pi$, $\rho$ and $\omega$: we must keep in mind their
momentum space behaviour, i.e., pseudoscalar (P) or vector (V) character
and the isospin (0 or 1);
\item we must also account for the number of $\Delta$'s in the
internal lines: $n_\Delta$ = 0, 1 or 2.
\end{enumerate}
Then the traces can be factorized as
$$\lambda T^{I}_{ij} S_{\mu\nu}$$
where $S$ results from the spin part and reads
\begin{eqnarray}
S_{PP}&=&4\left[{\bf q}\cdot \left({\bf q}+{\bf
k}\right)\right]^2\\
S_{PV}=S_{VP}=S_{VV}&=&4\left[q^2 k^2-
\left({\bf q}\cdot{\bf k}\right)^2\right]
\end{eqnarray}
$T$ comes  from the isospin traces and holds
\begin{eqnarray}
T_{00}^{T=0}=4& \qquad\qquad&T_{00}^{T=1}=0\\
T_{01}^{T=0}=T_{10}^{T=0}=0& \qquad\qquad&T_{01}^{T=1}=
T_{10}^{T=1}=4\\
T_{11}^{T=0}=12& \qquad\qquad &T_{11}^{T=1}=\mp 8\\
\end{eqnarray}
where the sign in $T_{11}^{T=1}$ is minus for the direct diagrams
and plus for the exchange ones; finally $\lambda$ depends upon
the number of $\Delta$'s:
\begin{eqnarray}
\lambda^{T=0}&=&\left({4\over 9}\right)^{n_\Delta}\\
\lambda^{T=1}&=&\left(-{2\over 9}\right)^{n_\Delta}\,.
\end{eqnarray}
Finally we note that in the diagrams f) to k) an $\omega$ cannot
be exchanged: this reduces to 51 the number of diagrams
effectively needed.

\section{Results}\label{sect5}

In this section we present the results of our calculation for the
longitudinal response, which, in the infinite nuclear matter limit,
is  related to the polarization propagator
\begin{equation} \Pi^{\mu\nu}(x,y)=
<\Psi_0|T\left\{j^\mu(x),j^\nu(y)\right\}|\Psi_0> \label{lr1a}
\end{equation}
by
\begin{eqnarray} R_L(k,\omega)&=&-{1\over \pi}\,{Z\over
\rho}G_{E\,p}(k)\Pi^{00}(k,\omega) \label{lr2a}\\
&=&-{1\over \pi}\,{Z\over \rho}
G_{E\,p}(k)\int  d(x-y)e^{ik\cdot(x-y)}\Pi^{00}(x-y),\nonumber
\end{eqnarray}
$\rho$ being the nuclear density.

We have shown in the previous sections that, with the help of the
analytical  results from Refs.~\cite{CeSa-88,CeCoCoSa-93-t}, the
computational  problem reduces to the  evaluation of one three-dimensional
integral for each diagram; the numerical integration
remains, however,  highly delicate because the functions
$\Pi^{(n)}$  exhibit a large number of singularities. Moreover, the
large predicted cancellations between different terms require a
great numerical precision.

The parameters used in the calculation are, unless otherwise
specified,  $g^\prime_0=0.5$,  $q_{c\,L}=800$ MeV/c,
$q_{c\,T}=1100$ Mev/c and $k_F=1.36~{\rm fm}^{-1}$;  we
compare the results with the data of Meziani et al
\cite{Me84,Me85} on  $Ca^{40}$ and $C^{12}$.

In fig. \ref{data1} we present the pion contribution to diagram a),
singling out the first order term from the remaining part of the
RPA  series, to stress the role of RPA-dressing of the internal line;
in fig. \ref{data2} we plot the total for diagram a), as well as the
contributions from the different mesons. The relevance of the
$\rho$ -- or, more correctly, of the effective interaction
in the isovector spin-transverse channel -- is  evident: this feature
persists in the  other diagrams.

Next we come to the self-energy diagrams  [b) and c)]. In fig.
\ref{data3} we present the pion part, singling out first  order and
RPA corrections both for diagrams b) and c); we remark that a
large cancellation occurs between them (they should cancel
exactly in the case of a constant  potential). In fig. \ref{data4} we
display the total of b) and c), again singling out  the terms from
different mesons.

Concerning the self-energy diagrams a peculiar difficulty arises,
namely the  instability of the calculation at the edges of the
response region; the reason is that these terms contain a
self-energy insertion on the fermionic lines: but this is only the
first term of a Dyson series. We know 'ab initio' that summing the
whole series would correspond to a modification of the dispersion
relation for the nucleon and, consequently, of the response region
itself. It is not surprising then that the inclusion of the first term
of the series only gives rise to wild oscillations: the convergence of
the loop expansion is warranted by the smallness of the 1-loop
contribution with respect to the mean field one: this is clearly not
the case just around the edges of the response region.
On the other hand the inclusion of the whole Dyson series for the
nucleons is inconsistent with the loop expansion scheme and it
should lead to the violation of sum rules and  general theorems.

The sharpness of the Fermi surface is responsible for a large part
of the problem:
from the mathematical point of view, in fact, the diagram
diverges when a pole in the denominator comes to the
boundary of the  integration region, and this happens just because the
Fermi surface is sharp: this is not the case for a finite nucleus,
thus the singularity could be  removed by smearing somewhat the
Fermi surface. We
stressed before that such a procedure would require an  enormous
increase of the computational difficulties, so we attributed the
nucleon a small  finite width at momenta near the Fermi surface,
to simulate the same effect. This correction is not included in  figs.
\ref{data3} and \ref{data4} but it will be accounted for in the
final result.

Next we come to diagrams d) and e). Here $\pi$, $\rho$ and
$\omega$ can contribute in all possible combinations.  The single
terms are shown in fig. \ref{data5}, where we have
plotted the sum of diagrams d) and e) separating the two pions,
two $\rho$'s, two $\omega$'s and finally
the mixed  ($\pi\rho$, $\pi\omega$ and $\rho\omega$)
contributions. Fig. \ref{data6} displays instead separately
diagrams d) and e). The effect of the $\Delta$'s in
the intermediate  states is illustrated in fig. \ref{data7}, where we
reported the total contributions  without,  with one and
two $\Delta$'s.  It is remarkable the strong cancellation occuring
between the contributions from diagrams b) and c), which would
strongly quench the response, and the ones from the other
diagrams featuring the opposite. Finally we present separately the
total from exchange, self-energy and  correlation diagrams in fig.
\ref{data8}.

We evaluated all the results presented until now always using the
set of parameters given at  the beginning of this chapter, with
the coupling constants and cutoffs of  sect. \ref{sect3}. Before
presenting the comparison between our  theoretical calculation
and the experimental data we examine the  sensitivity of
the results to small changes of the parameters. Keeping  the
coupling constants and cutoffs as fixed, we display in fig.
\ref{data9} the effect of variations of $g^\prime$, $k_F$ and of the
many-boy cutoffs $q_{c\,L,T}$.

Finally we compare with the experimental data in figs.
\ref{data10} and \ref{data11} for $^{40}Ca$ and $^{12}C$
respectively. In  the second case we assume $k_F=1.2~{\rm fm}^{-
1}$ to effectively take into account of the relevant smearing of the
surface in such a light nucleus.

\section{Discussions and outlook}\label{sect6}
In the previous section we presented the results of our
calculation with a given choice of the model parameters. Now we
need a  critical analysis of these outcomes.

As a first point, let us come back to our theoretical approach to
the  many-body problem. The loop expansion automatically entails the
fulfillment, order-by-order, of general theorems, sum-rules and so on. This
because the loop expansion is a, if case asymptotic, well-behaves series of
powers in a suitably chosen parameter, i.e., the coefficient multiplying
the action, to be identified at the end of the calculation with $\hbar$. It
is remarkable that, unlike the usual application of SPA, in our case the
formal parameter of the expansion does not simply coincide with $\hbar$,
because the Hubbard-Stratonovich transformation makes the effective bosonic
action itself $\hbar$ dependent. This implies, in turn, that if the value
of a given sum-rule does not depend on $\hbar$, then this sum-rule must be
fulfilled at the mean field level: contributions coming from higher orders
must separately vanish. This is the case, for instance, for the usual $f$
sum-rule, which reads:
\begin{equation}
\int_0^\infty\,d\omega\,R_L^{T=0}(q,\omega)\,=\,\frac{q^2}{2M}
\label{fsum}
\end{equation}
if in$R_L^{T=0}(q,\omega)$ the electromagnetic nucleonic form factors are
omitted. In fact the FFG longitudinal response fulfills eq.~(\ref{fsum});
the contributions from quantum fluctuations must then vanish, thus
realizing explicitly the expected cancellations between Feynman diagrams at
the same order. We verified the $f$ sum-rule in order to check the
numerical accuracy and the inner consistency of our approach. The task is
complicated because this sum-rule is energy weighted, i.e., contributions
of the high-energy tale is emphasized. This requires, first of all, to
extend the integration well beyond the QEP region.

We considered the two extreme cases $q=300$ and $410~MeV/c$ and then we
extended the integration up to $\omega=400~MeV$, cfr.~Figs.~15-19,
obtaining a saturation of the sum-rule of the order of $90\% $ in both
cases. To be more specific, at $q=300~MeV/c$, the exact value of the
sum-rule is $q^2/2M=47.97~MeV$: with the integral extended up to
$\omega=200~MeV$ -- containing anyway the whole region of the QEP -- we
obtain from l.h.s. of eq.~(\ref{fsum}) a value of $39.16~MeV$, while
extending the integral up to $\omega=400~MeV$ we get a sum-rule value of
$42.39~MeV$. At $q=410~MeV/c$, extending the integral up to
$\omega=400~MeV$, we obtain a value of $80.47~MeV$ to be compared with
$q^2/2M=89.60~MeV$. We consider this result quite satisfactory because it
is clearly beliavable that the remaining contribution from the tale will
saturate completely the $f$ sum-rule. We remark anyway that we estimate a
numerical uncertainty of the calculation of the order of $2\% $.

Coming to the detailed shape of the response,
we observe that, {\em a posteriori}, the
one-loop  corrections to the mean field results are sizable, but still
remarkably  smaller than the mean field itself; this is not a trivial
point,  because the mesons included in the model strongly interact
with the  nuclear matter: then very large corrections, mainly for
the case of the  $\rho$-exchange, could well be expected -- and
this is the case indeed for some diagrams. But large cancellations
among them are present and the final answers are  reasonably
small, indicating that the loop expansion works properly at  this
order. We expect, in fact, the loop expansion to be an asymptotic
series and only an explicit evaluation to a given order can establish if it
is still in the convergence region or not. Of course even an estimate of
the two-loop corrections seems presently too hard to be performed.

The previous statement, i.e., that the approximation scheme works
well at the  one-loop order, requires however to be somehow
weakened, because, as  already remarked, we expect a failure of
the first order  calculations near the edges of the QEP: in fact our
approach is not  able to provide a shift of the whole peak, which is
instead induced by  the dressing of the nucleon in the medium, i.e.
by rewriting the nucleon  propagator as
\begin{equation} {1\over p_0-{p^2\over 2m}\pm i\eta}\to
{1\over p_0-{p^2\over 2m} -  \Sigma(p,p_0)}\;.
\label{uuu}
\end{equation}
Our scheme contains, in fact, diagrams with one self-energy insertion
[diagrams b) and c)], but the whole Dyson's series should be summed up to
shift the position of QEP. Thus we expect to account for the
strength of the  peak but
\begin{enumerate}
\item the position of the peak should remain around the free one;
\item at the border of the QEP almost all the response is
generated by  the infinite series of self-energy insertions summed
up through the Dyson's  equations. Since only the first term of this
series is present, at the edges of the FFG response region the loop
expansion necessarily  fails.
\end{enumerate}
Just to give the reader a visual feeling of how a simple shift could
affect the response, we have reported in fig. \ref{data13} one of the
results of fig. \ref{data11}, but shifting the curve of 30
MeV upward. With this (formally unjustified) trick the agreement
with the data  becomes impressive, opening an interesting
question: clearly the corrections stemming from the loop
expansion resonably describe, in a channel dependent way, the
photoexcitation mechanism. This was our main purpose, but on
the other hand this approximation fails to describe the mechanism
leading to the nuclear binding, as it is clearly shown in fig.
\ref{data13}. This should be expected, but leaves the question:
``Could someone put all the things together?''

What discussed above concerned the theoretical framework. Now
we come to the dynamical model employed in  the present work.
This topic is not so well grounded on formal  properties of field
theory, but of course is much more strictly linked  with the
physical domain: the model, as it comes out
clearly from our discussion, is not parameter-free. Many
parameters are not experimentally well determined. Our  choice
was to keep the coupling constants and the cut-off as near as
possible to commonly accepted values, using the Bonn potential as
guideline for those quantities for which we have no sufficiently
clear  experimental evidence. In this way the number of
parameters is  drastically reduced. Even $g^\prime$ can vary only
between narrow limits  (i.e. $g^\prime=0.5\div 0.6$ in non static
conditions \cite{CeSa90}) and for $k_F$ we made the choice
$k_F=1.36~{\rm fm}^{-1}$ for $^{40}Ca$  and $k_F=1.2~{\rm fm}^{-
1}$ for $^{12}C$, which seems quite natural.

The many-body cut-off $q_{cL}$ can also vary, reasonably,
between 800  and 900 MeV/c, while, finally, the cut-off on the
spin-transverse  channel has been treated as an approximately
free parameter. Between these really narrow constraints we have
chosen the most favourable situation to reproduce the experimental
data of \cite{Me85}.

We observe that the choice of $k_F$ and $g^\prime$ seems to be
quite  reasonable, while, according to the discussion of sect.
\ref{sect3} a  rather higher value for $q_{cL}$ (like for instance
$q_{cL} = 900$  MeV/c) should be preferable.  The net effect
should be a small reduction of the quenching of the  quasi elastic
peak.

Very recently a new analysis of the Rosenbluth separation
\cite{YatWi93}, based on experimental data taken at the Bates
Laboratory,  seemed to indicate a reduction of the longitudinal
response quite lower  than the one previously believed. The
discrepancies between the two  results indicate that more
careful experimental analyses are required. It is of course outside
our  scope to discuss the validity of the experimental data: we
only present  here a comparison of the two sets of results,
together with our  theoretical outcomes obtained with two
different sets of parameters, namely $g'=0.5$, $q_{cL}=800$ MeV/c,
$q_{cT}=1100$ MeV/c and $g'=0.6$, $q_{cL}=900$ MeV/c,  $q_{cT}=1200$
MeV/c (see fig. \ref{data14}).  The first set better approaches the
results of  Saclay, while the second is more suitable to reproduce
the Bates data, at least in the left side of the peak.

To summarize we believe to have shown that, within a mesonic
theory, a 1--loop calculation provides a channel dependent
photo-excitation mechanism able to significantly quench the longitudinal
response. The size of this quenching is, however, not completely
determined, both theoretically and experimentally. In the present model the
major uncertainty comes from the effective interaction in the
$\rho$--channel: combined changes in $g'$, $q_{c\,T}$ and $\Lambda_\rho$
can significantly affect the response - without modifying, however the
trend illustrated so far. Exclusive (e$\rm e'$,2N) experiments are
required to improve our understanding of the p-h effective
interaction\cite{Ce-93-t} and to better fix the parameters of the model.
Coming to the perspectives opened by the present approach,
the next step will be, obviously, to extend the
calculation to the transverse (magnetic) response making use of
the same theoretical scheme: the number of diagrams to be
evaluated increases rapidly, because both the MEC and the direct
photo-excitation of the $\Delta$ are present.
In particular, the possible electromagnetic excitation of $\Delta$-h pairs
can be viewed as a channel dependent vertex renormalization in the medium
-- thus coming back to the idea of M. Ericson and M. Rosa-Clot\cite{ErRo-86}
-- and could be well responsible for the different behaviour of the
longitudinal and transverse channel. Remarkably this feature emerges at the
1--loop level. Unfortunately,
some new parameters -- the meson-$\Delta\Delta$ couplings --
will enter the game. This calculation is presently in progress.

More response functions can be evaluated along these lines, like,
e.g., the isoscalar-spin longitudinal and transverse ones\cite{Peterson}.
Here an interesting opportunity is offered, because direct $\Delta$
excitation is forbidden in both channels. Thus, if our guess about the
relevance of the $\Delta$ excitation in simultaneously explaining the
electromagnetic responses is correct, the separation of the spin
logitudinal and transverse responses in the $S=1$,
$T=0$ channel should give rise to less pronounced differences.

Furthermore, the same model can be applied to the
parity-violating responses\cite{Donnelly}, where a good confidence in the
microscopical model is needed to establish the feasibility of experiments
able to detect the $\gamma-Z_0$ interference in the electron scattering.

Moreover, as far as higher momentum data will become available,
other issues will have increasing relevance: first of all the study of
relativistic effects, at least those coming the relativistic kinematics
of the nucleons.

\section*{Figure Caption}
\begin{figure}[ht]
\caption{\protect\label{fig1}One loop diagrams. (a) exchange diagram, (b)
and (c): self-energy diagrams, (d) and (e): correlation diagrams.}
\end{figure}
\begin{figure}[ht]
\caption{\protect\label{fig2}Diagrams inducing SRC on the pion
exchange.}
\end{figure}
\begin{figure}[ht]
\caption{\protect\label{fig3}Diagrams with overcounting in two
successive correlated pion exchanges.}
\end{figure}
\begin{figure}[ht]
\caption{\protect\label{fig4cd}
Diagrams (f), (g), (h) and (i) [see text] with one $\Delta$ line.}
\end{figure}
\begin{figure}[ht]
\caption{\protect\label{fig4e}Diagrams (j) and (k) [see text]
with two $\Delta$ lines.}
\end{figure}
\begin{figure}[ht]
\caption{\protect\label{data1}
Contribution to $R_L$ of the $\pi$-exchange, diag. a), for $^{40}Ca$
at $q=300~MeV/c$. Dotted line:
first order, dashed lines: remaning part of RPA, solid line: total.}
\end{figure}
\begin{figure}[ht]
\caption{\protect\label{data2}
Contribution to $R_L$ of the $\pi$, $\rho$ and $\omega$-exchanges,
diag. a), for $^{40}Ca$ at $q=300~MeV/c$. Dotted line:
$\pi$, dashed line: $\rho$, solid line: $\omega$.}
\end{figure}
\begin{figure}[ht]
\caption{\protect\label{data3}
Contribution to $R_L$ of the $\pi$-exchange, diag. b) and c), for $^{40}Ca$
at $q=300~MeV/c$.
Dotted line: first order diag. b), thick dots: first order
diag. c); dashed line: RPA corrections diag. b),dash-dotted line: RPA
corrections diag.c); solid line: total.}
\end{figure}
\begin{figure}
\caption{\protect\label{data4}
Contribution to $R_L$ of the $\pi$, $\rho$ and $\omega$-exchanges,
diag. b) and c), for $^{40}Ca$ at $q=300~MeV/c$.
Pion: dotted line; $\rho$: dashed line, $\omega$:
dash-dotted line; total: solid line.}
\end{figure}
\begin{figure}
\caption{\protect\label{data5}
Contribution to $R_L$ from diags. d) and e) for $^{40}Ca$ at $q=300~MeV/c$.
Dots: $\pi\pi$,
dashes: $\rho\rho$, dash-dots: $\omega\omega$, dots with stars:
$\pi\rho$, dashes with stars: $\pi\omega$, dash-dots with stars:
$\rho\omega$; solid: total.}
\end{figure}
\begin{figure}
\caption{\protect\label{data6}
Contribution to $R_L$ from diags. d) and e) for $^{40}Ca$ at $q=300~MeV/c$.
diagrams d): dashed line; diagrams e): dash-dotted line; total: solid
line.}
\end{figure}
\begin{figure}
\caption{\protect\label{data7}
Contribution to $R_L$ from diags. d) to k) for $^{40}Ca$ at $q=300~MeV/c$.
no $\Delta$'s, diags. d) and e): dashed line;
one intermediate $\Delta$, diagrams f), g), h) and i): dash-dotted line;
two intermediate $\Delta$'s, diagrams j) and k): dotted line;
total: solid line.}
\end{figure}
\begin{figure}
\caption{\protect\label{data8}
Contribution to $R_L$ from diags. a) to k) for $^{40}Ca$ at $q=300~MeV/c$.
Contribution of exchange diagram a):
dashed line; self-energy diagrams b and c): dash-dotted line;
correlation diagrams d) to k): dotted line; total: solid line.}
\end{figure}
\begin{figure}
\caption{\protect\label{data9} Sensitivity of the one-loop contribution
to $R_L$ to the many-body parameters. All parameters but the ones explicitly
indicated are those previously stated. Solid line: standard parameters;
dashed line: $g^\prime=0.6$, dotted line: $k_F=1.2 {\rm fm}^{-1}$,
dash-dotted line: $q_L=900$ MeV/c, $q_T=1200$ MeV/c.$R_L$ for $^{40}$Ca
is displayed
in MeV$^{-1}\times$1000 versus the
transferred energy (in MeV).}
\end{figure}
\begin{figure}
\caption{\protect\label{data10} Comparison of $R_L$
with experimental data [taken
from ref. \protect\cite{Me85}] for $^{40}Ca$ as a function of the energy (in
MeV). The dashed line refers to the Free
fermi gas contibution, the solid line represents the total results,
while in the dash-dotted line the contribution of the diags. without
internal $\Delta$ lines [diags. f) to k)].}
\end{figure}
\begin{figure}
\caption{\protect\label{data11} As fig. \protect\ref{data10} but data are taken
from $^{12}C$ and here $k_F=1.2~{\rm fm}^{-1}$.}
\end{figure}
\begin{figure}
\caption{\protect\label{data13} As fig. \protect\ref{data10}, but the whole
curve has been shifted [see text for explanation]. Only the case
$q=300~MeV/c$ is plotted.}
\end{figure}
\begin{figure}
\caption{\protect\label{data14} Response functions compared with the
data of ref. \protect\cite{Me85} and \protect\cite{YatWi93}.
Solid line: results with $q_{cL}=800$ MeV/c and $q_{cT}=1100$ MeV/c;
dashed line: results with $q_{cL}=900$ MeV/c and $q_{cT}=1200$ MeV/c.}
\end{figure}

\section{Tables}

\begin{table}[ht]
\begin{tabular}{|c|c|c|c|}
Particle&Channel&Diag. a)&Diag. b)\\
{}~&$T=0$&  12 $q^2$  &  $12q^2  $\\
\cline{2-4}
$\pi$&T=1&  $-4q^2  $&  $12q^2  $\\
\cline{2-4}
{}~&e.m.&  $2q^2 $ &  $6q^2  $\\
{}~&$T=0$&  $24q^2  $&$  24q^2  $\\
\cline{2-4}
$\rho$&$T=1$&  $-8q^2  $&  $24q^2  $\\
\cline{2-4}
{}~&e.m.&  $4q^2  $&  $12q^2  $\\
{}~&$T=0$&  $4q^2  $&  $8q^2  $\\
\cline{2-4}
$\omega$&$T=1$&  $4q^2  $&  $8q^2  $\\
\cline{2-4}
{}~&e.m.&  $2q^2  $&  $4q^2  $\\
\end{tabular}
\caption{\protect\label{tab1}Table of the spin-isospin traces for
the
diagrams a) and b) (diagram c) coincides with b))}
\end{table}

\end{document}